\documentclass[sigconf]{acmart}

\usepackage{CJKutf8}
\usepackage{xcolor}
 \usepackage{float}
 \usepackage{amsmath}
 \usepackage{array}
\usepackage{enumitem}
\newlist{subquestion}{enumerate}{1}
\setlist[subquestion,1]{label=(\alph*)}

\newcommand{\sysName}{\textit{InsightMap }}
\newcommand{\sysNameWithoutSpace}{\textit{InsightMap}}

\newcommand{\squishlist}{
 \begin{list}{$\bullet$}
 { \setlength{\itemsep}{0pt}
   \setlength{\parsep}{3pt}
   \setlength{\topsep}{3pt}
   \setlength{\partopsep}{0pt}
   \setlength{\leftmargin}{1.2em}
   \setlength{\labelwidth}{1em}
   \setlength{\labelsep}{0.6em}
 }
}
\newcommand{\squishend}{
 \end{list}
}

\newcommand{\squishlistCat}{
 \begin{list}{$\bullet$}
 { \setlength{\itemsep}{0pt}
   \setlength{\parsep}{3pt}
   \setlength{\topsep}{3pt}
   \setlength{\partopsep}{0pt}
   \setlength{\leftmargin}{2em}
   \setlength{\labelwidth}{1.1em}
   \setlength{\labelsep}{0.6em}
 }
}
\newcommand{\squishCatend}{
 \end{list}
}

\newcommand{\stitle}[1]{\vspace*{0.4em}\noindent{\bf #1.\/}}

\AtBeginDocument{%
  }

\setcopyright{acmlicensed}
\copyrightyear{2025}
\acmYear{2025}
\setcopyright{rightsretained}
\acmConference[CHI EA '25]{Extended Abstracts of the CHI Conference on Human Factors in Computing Systems}{April 26-May 1, 2025}{Yokohama, Japan}
\acmBooktitle{Extended Abstracts of the CHI Conference on Human Factors in Computing Systems (CHI EA '25), April 26-May 1, 2025, Yokohama, Japan}\acmDOI{10.1145/3706599.3719702}

\acmISBN{979-8-4007-1395-8/2025/04}

\begin{document}

\title[\sysNameWithoutSpace]{Data Insights as Data: Quick Overview and Exploration of Automated Data Insights}

% \author{Shangxuan Wu$^{1}$, Wendi Luan$^{1}$, Yong Wang$^{2*}$, Dan Zeng$^{3}$, Qiaomu Shen$^{1*}$, Bo Tang$^{1}$}
% % \authornote{Both authors contributed equally to this research.}
% \affiliation{
%   Southern University of Science and Technology$^1$;
%   Nanyang Technological University$^2$;\\
%   Sun Yat-sen University, School of Artificial Intelligence$^3$
%   \\
%   \textit{\{wusx2018,11712532\}@mail.sustech.edu.cn};
%   \textit{yong-wang@ntu.edu.sg};\\
%   \textit{zengd8@mail.sysu.edu.cn};
%   \textit{\{shenqm,tangb3\}@mail.sustech.edu.cn};
%   \institution{}
%   \city{}
%   \state{}
%   \country{}
% }

\author{Shangxuan Wu}
\affiliation{
  \department{Department of Computer Science and Engineering}
  \institution{Southern University of Science and Technology}
  \city{Shenzhen}
  \state{Guangdong}
  \country{China}
  }
\email{wusx2018@mail.sustech.edu.cn}

\author{Wendi Luan}
\affiliation{
\department{Department of Computer Science and Engineering}
  \institution{Southern University of Science and Technology}
  \city{Shenzhen}
  \state{Guangdong}
  \country{China}
  }
\email{11712532@mail.sustech.edu.cn}

\author{Yong Wang}
\authornote{Qiaomu Shen and Yong Wang are the corresponding authors.}
\affiliation{
\department{College of Computing and Data Science}
  \institution{Nanyang Technological University}
  \city{Singapore}
  \state{Singapore}
  \country{Singapore}
  }
\email{yong-wang@ntu.edu.sg}

\author{Dan Zeng}
\affiliation{
  \department{School of Artificial Intelligence}
  \institution{Sun Yat-sen University}
  \city{Zhuhai}
  \state{Guangdong}
  \country{China}
  }
\email{zengd8@mail.sysu.edu.cn}

\author{Qiaomu Shen}
\authornotemark[1]
\affiliation{
  \department{Department of Computer Science and Engineering}
  \institution{Southern University of Science and Technology}
  \city{Shenzhen}
  \state{Guangdong}
  \country{China}
  }
\email{shenqm@sustech.edu.cn}

\author{Bo Tang}
\affiliation{
  \department{Department of Computer Science and Engineering}
  \institution{Southern University of Science and Technology}
  \city{Shenzhen}
  \state{Guangdong}
  \country{China}
  }
\email{tangb3@sustech.edu.cn}

\renewcommand{\shortauthors}{Shangxuan Wu, Wendi Luan et al.}

\begin{abstract}
Automated data insight mining and visualization have been widely used in various business intelligence applications (e.g., market analysis and product promotion).
However, automated insight mining techniques often output the same
mining results to different analysts without considering their personal preferences, while interactive insight discovery requires significant manual effort.
This paper fills the gap by integrating automated insight mining with interactive data visualization and striking a proper balance between them to facilitate insight discovery and exploration.
Specifically, we regard data insights as a special type of data and further present \sysNameWithoutSpace, a novel visualization approach
that uses the map metaphor to
provide a quick overview and in-depth exploration of different data insights, where a metric is proposed to measure the similarity between different insights.
The effectiveness and usability of \sysName are demonstrated through extensive case studies and in-depth user interviews.
\end{abstract}




\maketitle

% \maketitle
\section{Introduction}

Insight discovery and exploration are essential tasks in data analytics across various applications, often achieved through data visualization \cite{card1999readings}. 
Analysts interact with visualizations to perform exploratory data analysis, but this process can be tedious and time-consuming, especially for large and complex datasets. 
Alternatively, statistical modeling and machine learning techniques have been used to automatically extract insights \cite{ding2019quickinsights,vartak2015seedb,tang2017extracting}. 
While these methods improve efficiency, they lack contextual awareness to determine the most relevant insights for different analysts \cite{mcnutt2020divining}. 
Moreover, their outputs are typically presented as isolated insights without a comprehensive overview, limiting further analysis and critical thinking. 
For example, \textit{
what is the overall distribution of data insights across different insight types and subsets of the whole dataset?
What are the relations between different data insights? 
}
It is crucial to provide users with \textbf{a quick overview of insight distribution} within a dataset and \textbf{a detailed exploration of the relations between different data insights}.

However, achieving this analytical goal presents significant challenges, primarily from two perspectives.
First, a given dataset may contain a vast number of diverse data insights, making it nontrivial to provide analysts with a concise yet comprehensive overview that effectively summarizes all insights. Without such an overview, analysts may struggle to navigate and interpret the full scope of insights within the dataset.
Second, each data insight is inherently multidimensional, involving various attributes such as its type, associated data items, and key attributes. 
Also, insights are often interconnected, exhibiting complex relationships across different dimensions, including subspace similarities and insight categories.
It remains unclear how to characterize different data insights in a generic and quantitative way, as well as how to further measure the relations among data insights and dataset subspaces.

To address this research gap, we propose \textit{regarding data insights as data} and develop {\sysNameWithoutSpace}, a visual analytics system that integrates automated analysis with a map-based visualization for effective insight exploration. 
We first extract diverse data insights (e.g., outlier, correlation, trend) of a multi-dimensional dataset by using QuickInsights \cite{ding2019quickinsights}. 
To quantify insight similarity, we employ a similarity metric commonly used for categorical data.
Building on this, we design a visual analytics system with three linked views:
\textit{Data Distribution View}  provides an overview of the multi-dimensional dataset.
\textit{Insight Map View} visualizes insight distribution and relationships using a map metaphor, allowing insights to be grouped into quadrants based on user preferences.
\textit{Individual Insight View} presents each insight with intuitive charts and brief text descriptions.
We validate {\sysNameWithoutSpace} through a real-world case study and expert user interviews, demonstrating its effectiveness and usability across different domains. 
{\sysNameWithoutSpace} is open-source at https://github.com/ArslanaWu/InsightMap.

\section{Related Work}
The related work includes automated insight mining, insight-based visualization recommendation and exploratory visual analysis. 

\stitle{Automated Insight Mining}
Data mining techniques facilitate insight discovery by evaluating the interestingness of patterns using rule-based methods \cite{drosou2013ymaldb,chen2002multi,geng2006interestingness} and machine learning approaches \cite{Sun2023erato}. Microsoft researchers \cite{tang2017extracting,ding2019quickinsights} proposed a general framework for identifying insightful patterns based on impact and significance. Efficiency in automated mining has been improved through subspace pruning \cite{wu2009promotion}, further optimized in recent methods \cite{tang2017extracting,ding2019quickinsights}. Our approach follows \cite{ding2019quickinsights} to efficiently mine diverse data insights.

\stitle{Insight-based Visualization Recommendation}
Visualization recommendation research focuses on generating effective visualizations for exploration, emphasizing either perception-effective encodings \cite{hu2019vizml,dibia2019data2vis} or insightful data patterns \cite{demiralp2017foresight,cui2019datasite}. Recent work has improved logical organization and readability of visual insights \cite{shi2020calliope,mcnutt2020divining,chen2024calliopenet}. However, limited insight types and lack of interactivity often hinder deeper exploration \cite{li2021exploring}. To address this, \sysName employs a map metaphor to visually represent data insight distribution and relationships.

\stitle{Exploratory Visual Analysis (EVA)}
EVA supports discovering new insights beyond predefined goals \cite{battle2019characterizing,tukey1977exploratory,keim2006challenges}. Users may adopt bottom-up exploration for open-ended discovery \cite{liu2014effects,reda2016modeling,alspaugh2018futzing,zgraggen2016progressive} or top-down strategies with explicit hypotheses \cite{gotz2009characterizing,siddiqui2016effortless,zgraggen2018investigating}. Many combine both approaches \cite{reda2016modeling,lam2017bridging}. \sysName facilitates EVA by visualizing auto-extracted insights, enabling both overview and detailed analysis while supporting mixed exploration strategies.

\section{Design Considerations}
\label{sec:design-considerations}
% As pointed out by Heer~\cite{heer2019agency}, it is crucial to strike a proper balance between manual exploration and automated algorithms for data insight discoveries.
% Guided by this consideration, \sysName aims to help analysts gain a quick overview and detailed exploration of data insights extracted by an automated approach.
% When designing \sysName, we need to consider many design choices 
To ensure that \sysName effectively facilitates insight exploration,  
% To guide our designs of \sysName, 
we surveyed existing research on insight mining~\cite{sarawagi1998discovery,drosou2013ymaldb,tang2017extracting,ding2019quickinsights}, visualization recommendation~\cite{cui2019datasite,wongsuphasawat2015voyager,wongsuphasawat2017voyager,shi2020calliope,mcnutt2020divining}, principles for visualization design~\cite{shneiderman2003eyes,munzner2014visualization} and exploratory search~\cite{battle2019characterizing,tukey1977exploratory,keim2006challenges}, and further compile a list of design considerations after multiple iterations of discussions and designs. The major considerations include:

% \begin{enumerate}[label=\textbf{C{\arabic*}},nolistsep]
\squishlistCat
\item[C1]
\textbf{Provide an overview of data distribution.}
\label{C-overview}
Providing an overview of the dataset, including a summary of its dimensions and ranges, helps analysts quickly understand new data and identify insights of interest.

\item[C2]
\textbf{Display the overall distribution of insight similarities.}
"Overview First and Details on Demand" \cite{shneiderman2003eyes}, an overview helps analysts identify points of interest and related insights.

\item[C3]
\textbf{Show insight details on demand.}
After reviewing the data and insight distributions, analysts may zoom in on specific insights, requiring detailed information for full understanding.

\item[C4]
\textbf{Combine visuals and text for easy understanding.}
Intuitive visualizations help analysts quickly grasp visual encoding, while natural language descriptions explicitly convey insights~\cite{shi2020calliope,ding2019quickinsights,srinivasan2018augmenting}.

\item[C5]
\textbf{Enable Interactive steering to facilitate exploring insights of user interest.}
Analysts may have diverse and evolving interests in the same dataset, making rich interactions in \sysName essential.
% Different analysts may have varying interests in the same dataset, and these interests may evolve as they explore the data.
% For the same dataset, different analysts can have significantly-different interests. Also, analysts' interest may evolve as they browse their data.
% Thus, it is necessary to provide rich interactions in \sysName, so that analysts can steer automatically-generated data insights and check those insights of their interest deeply.
% It is necessary to provide rich interactions in \sysName.
% \end{enumerate}
\squishCatend

% \squishlistCat
% \squishCatend
\section{System Overview}

\begin{figure}[]
    % \vspace{5mm}
    \centering
    \includegraphics[width=1\linewidth]{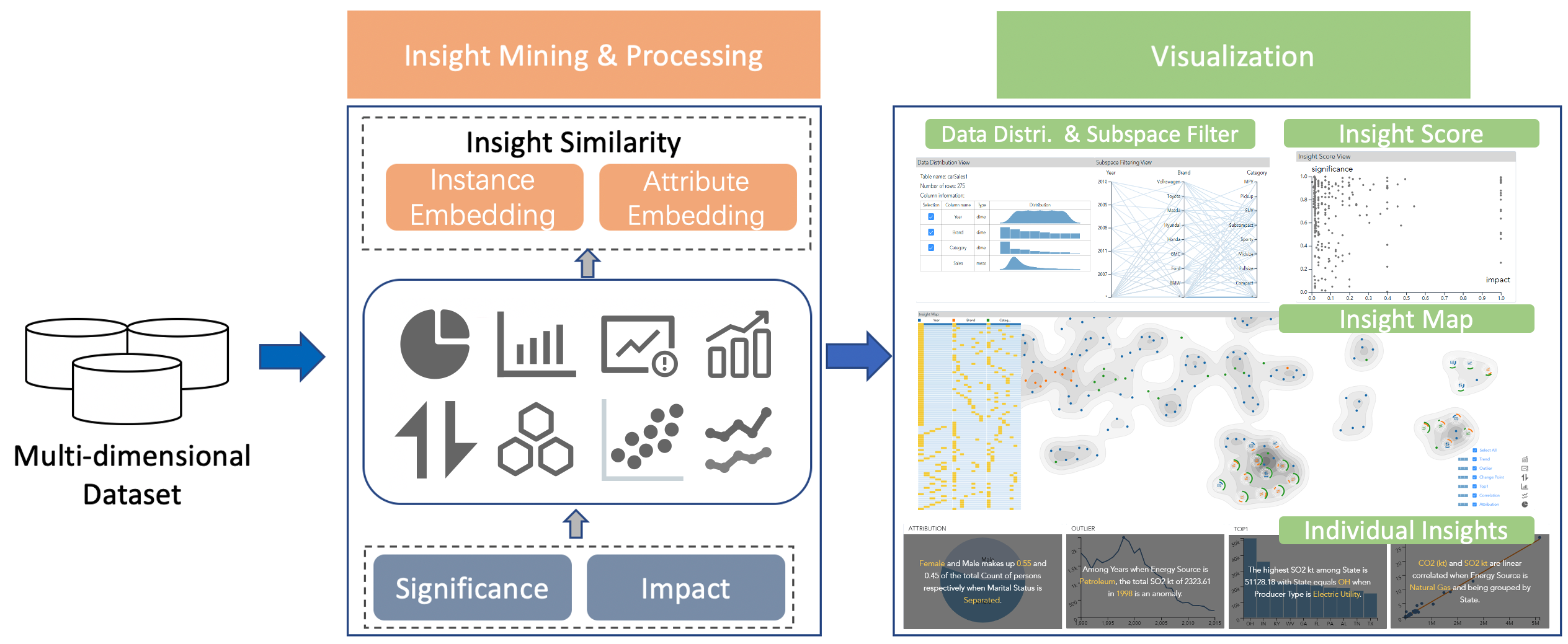}
    \caption{\sysName consists of two major components: insight mining and visualization. 
    The insight mining module extracts 8 types of different insights from the multi-dimensional dataset, while the  visualization module provides an integrated view of overall distribution and details.
    % overall distribution, subspace filtering, insight score distribution, insight map and individual insight views.
    % , enabling quick overview and detailed exploration of automated data insights. 
    }
    \label{fig_framework}
    % \vspace{-10mm}
\end{figure}

% Guided by the above design considerations, we proposed {\sysNameWithoutSpace} to integrate automated insight extraction with interactive visual analysis and enable quick overview and exploration of data insights.
% {\sysNameWithoutSpace} focuses on the insight analysis and exploration for multi-dimensional dataset, which are often stored as tables. 
% Each row is a data sample and each column represents a data attribute (also called a \textit{data field}). 
% Following the taxonomy of QuickInsights~\cite{ding2019quickinsights},
% we also classify data fields into two categories: \textit{dimensions} and \textit{measures}, where dimensions refer to categorical or ordinal attributes and measures refer to numerical attributes.

Guided by these design considerations, we developed {\sysNameWithoutSpace}, which integrates automated insight extraction with interactive visual analysis for efficient insight exploration. It focuses on multi-dimensional datasets typically stored as tables, where rows represent data samples and columns represent attributes (or \textit{data fields}). 
Following QuickInsights~\cite{ding2019quickinsights}, we classify data fields into two types: \textit{dimensions} (categorical or ordinal attributes) and \textit{measures} (numerical attributes).

% As shown in Fig.~\ref{fig_framework}, \sysName{} consists of two major modules: \textit{insight mining \& processing} and \textit{visualization}.
% The module of \textit{insight mining \& processing} employ QuickInsight~\cite{ding2019quickinsights}
% to extract 8 types of data insights from the multi-dimensional dataset, and further evaluate the similarity between any pair of data insights.
% A detailed list of the data insights will be introduced in Section~\ref{sec-insight-extraction}.
% The \textit{visualization} module allows analysts interact with the automatically extracted data insights. 
% As shown in Fig.~\ref{fig_UI}, it consists of three main views: \textit{Data Distribution View}, \textit{Insight Map View} and \textit{Individual Insight View}.
% The data distribution view provides analysts with a quick overview of the data distribution along each field. 
% The well-designed insight map view further leverages a map metaphor to show the overall distribution of data insights and the similarity among different insights,
% guiding analysts to explore insights of their interest.
% The insight details are visualized in an individual insight view for analysts to conduct in-depth checking, where a standard chart and a brief text description are shown to facilitate an easy understanding of each insight. 
% \sysName also provides rich interactions and enables linked analysis across different views.
As shown in Fig.\ref{fig_framework}, \sysName{} consists of two main modules: \textit{insight mining \& processing} and \textit{visualization}. 
The \textit{insight mining \& processing} employs QuickInsight\cite{ding2019quickinsights} to extract eight types of insights from a multi-dimensional dataset and evaluate their similarity (detailed in Section~\ref{sec-insight-extraction}). 
The visualization module enables interactive exploration through three views (Fig.~\ref{fig_UI}): \textit{Data Distribution View} for an overview of data distribution, \textit{Insight Map View} using a map metaphor to illustrate insight distribution and relationships, and \textit{Individual Insight View} for detailed examination with charts and descriptions. \sysName{} also supports rich interactions and linked analysis across views.

\section{Insight Mining and Processing}

\subsection{Insight Extraction}
\label{sec-insight-extraction}
We employ the recently proposed insight mining approach, QuickInsight~\cite{ding2019quickinsights}, to extract a series of data insights from the input dataset, as it can effectively detect representative data insights from the multi-dimensional dataset.
Also, it has been integrated into Microsoft Power BI~\footnote{https://docs.microsoft.com/en-us/power-bi/create-reports/service-insights} and is widely used by prior studies~\cite{mcnutt2020divining,shi2020calliope,srinivasan2018augmenting}.

\stitle{Insight Types} \label{sec:insight_type}
We extract eight types of insights from the input dataset using QuickInsights~\cite{ding2019quickinsights}, covering both single-field characteristics (i.e., attributes) and relationships between two fields. 
These insight types include: \textit{Top one}, \textit{Attribution}, \textit{Change point}, \textit{Outlier}, \textit{Trend}, \textit{Correlation}, \textit{Cross-measure correlation}, and \textit{Clustering}.
A comprehensive definition and extraction of different insights can be found in QuickInsight~\cite{ding2019quickinsights}.

\stitle{Insight Representation}
\label{sec-insight-representation}
Each data insight represents a piece of information that is automatically extracted from the dataset.
We follow the insight representation introduced in prior studies~\cite{ding2019quickinsights,shi2020calliope}, where each data insight can be represented as a 5-tuple: ${Insight} = \{Type, Subspace, Breakdown, Measure, Score\}$, 
where \textbf{Type} labels the insight category (Section~\ref{sec:insight_type}); \textbf{Subspace} is a set of filters that specify the data scope of an insight, which can be represented as ${Subspace} = \{ \{F_1 = V_1 \}, \{F_2 = V_2 \},..., \{F_k = V_k \} \}$,
where $F_k$ and $V_k$ represent a data field and its corresponding value for this subspace.
\textbf{Breakdown} is one or multiple categorical data field(s) that can further divide the subspace into different data groups.
\textbf{Measure} is a numerical data field where some aggregations such as \textit{average}, \textit{sum}, \textit{minimum}, \textit{count}, and \textit{gradient} can be performed on this subspace.
\textbf{Score} quantifies the overall ``interestingness'' of an insight, which indicates the importance and significance of each insight. A larger value of \textbf{Score} is preferred.

\subsection{Insight Similarity}

To provide analysts with a quick overview and enable effective exploration of data insights, we further propose assessing the similarity between each pair of insights. By measuring the similarity between different insights, analysts will be able to easily identify other insights that are potentially related to the current insight under their exploration.
As introduced in Section~\ref{sec-insight-representation}, an insight can be represented by five elements, where the most fundamental element is the subspace of an insight that delineates the data items related to the insight.
Thus, we choose to characterize the similarity between insights from the perspective of the insight subspace.
Specifically, we propose \textit{insight embedding vector} to depict the related data items (i.e., rows in a table) within the subspace of an insight.

\textbf{Insight Embedding Vector}:
Suppose the multi-dimensional dataset $D$ is denoted as $D = \{r_1, r_2, ... r_N\}$, where $N$ is the size of $D$ and $r_i$ represents the $i-$th row of the dataset.
For the raw dataset $D$, the value set of the $a-$th attribute is $y^a$, and the $k-$th value of $y^a$ is indicated by $y^{a, k}$. 
Moreover, for the raw data instance $r$, $r^{a}$ represents the value of $a-$th attribute.
With the given insight  ${insight_i} = \{type_i, subspace_i, breakdown_i, measure_i, score_i\}$, we define two categories of embedding vectors: \textbf{instance coverage} and \textbf{attribute coverage}. 
\squishlist
    \item \textbf{Instance coverage embedding}  of $insight_i$ is $ vecI_i = 
    < v_i^{1}, ... v_i^{N} >$, where $v_i^{j}$ equals $1$ if ${r_j} \in subspace_i$ otherwise $0$.
    \item \textbf{Attribute coverage embedding} of $insight_i$ is $vecA_i = <..., v_i^{a,k},...>$, where $v_i^{a, k} = |\{r| \forall r \in subspace_i \;and\; r^a=y^{a,k}\}| / N$,  is the attributes and $|\cdot|$  measures the cardinality of a set.
\squishend

\textbf{Insight Similarity Characterization}:
With the embedding vector, we can easily measure the similarity between insights by calculating the distance between two data insights in the high-dimensional space, such as the Euclidean Distance.
When projecting all the data insight embeddings into 2D plane with dimension reduction techniques, the overall similarity between different data insights can be easily recognized.

\begin{figure*}[h!]
    \centering
    \includegraphics[width=0.95\linewidth]{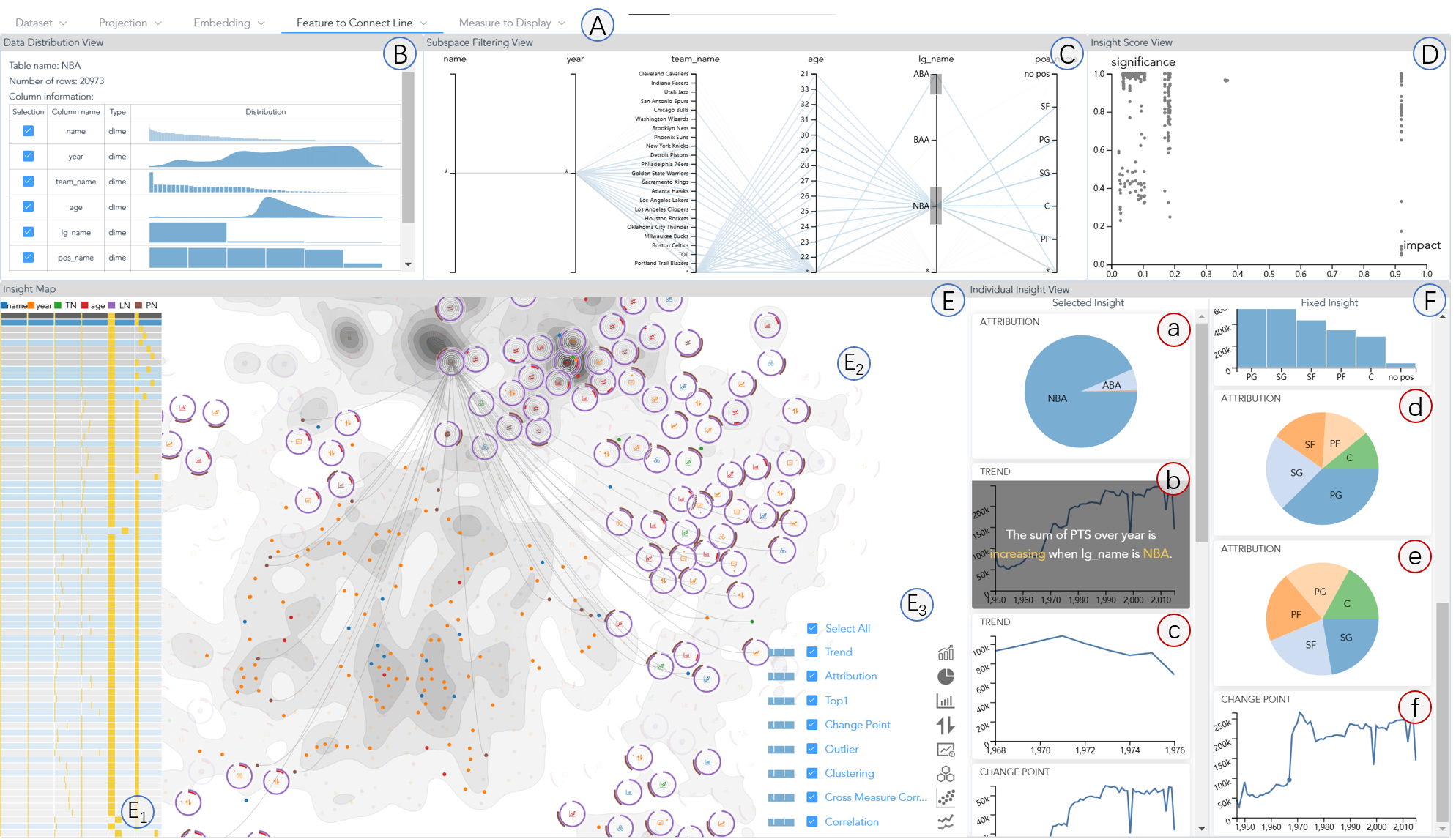}
    \caption{The user interface of \sysName. It consists of (A) \textit{Drop-down Menus} for parameter and visualization settings, (B) \textit{Data Distribution View} for overall attribute distributions, (C) \textit{Subspace Filtering View} for filtering subspaces via attribute axes, (D) \textit{Insight Score View} for insight significance and impact, (E) \textit{Insight Map View} for visualizing insights as glyphs and dots, and (F) \textit{Individual Insight View} for detailed insight inspection. This figure showcases automated data insights of the NBA dataset (Section~\ref{sec-case-study}).}
    \label{fig_UI}
    % \vspace{-4mm}
\end{figure*}

\section{Visualization}
% Guided by the design considerations introduced in Section~\ref{sec:design-considerations}, we have developed
% \sysName aims to help analysts quickly explore and analyze automated data insights through visualization. 

This section describes the major visual designs of \sysName that assist analysts in quickly exploring data insights.

\subsection{Data View}
% \wsxtodo{The section divisions are unreasonable and need to be revised}

% \subsection{Data Distribution View}
% According to \ref{C-overview}, it is necessary to provide analysts with an quick overview of the input data, which can give analysts a basic understanding on the data.
% Since the most important information of a multi-dimensional dataset is its data field names (i.e., column names), data types and the range of each data field, we propose using a table to briefly show such key information in a table (Fig.~\ref{fig_UI}B). More specifically, the data types can be \textit{dimension} or \textit{measure}, where dimension corresponds to categorical or ordinal values and measure indicates numerical values~\cite{ding2019quickinsights}. 
% The detailed distribution of data field is visualized by either histograms or area charts, displaying the number of data rows with each specific value at a data field. 
% Other important information for a dataset, such as it name and number of rows, is also shown in the data distribution view.
% With a glance at the data distribution view, analysts can quickly know the basic characteristics of the input data.
\stitle{Data Distribution View} To give analysts a basic understanding on the data, we use tables to display key information about multidimensional data, such as column names, data types and the range of each data field. The detailed distribution of data field is visualized by either histograms or area charts. With a glance at the \textit{Data Distribution View}, analysts can quickly know the basic characteristics of the input data.

\stitle{Subspace Filtering View} 
% To enable users to interactively filter and explore subspaces of interest, we designed the subspace filtering view. 
% As shown in Fig.~\ref{fig_UI}C, a parallel coordinates plot (PCP) displays the selected data fields and the distribution of subspaces associated with data insights. 
% Each PCP axis represents a data field and its possible values, while each line indicates a subspace linked to at least one insight. 
% The lowest tick, marked with an asterisk (*), signifies no restrictions on that field (i.e., the subspace can take any value). 
% Analysts can brush along axes to select fields and values of interest, filtering and highlighting relevant subspaces in the Insight Map View for quick exploration of targeted insights.
To enable interactive subspace filtering, we designed the \textit{Subspace Filtering View} (Fig.~\ref{fig_UI}C). 
It features a parallel coordinates plot (PCP) where each axis represents a data field and each line corresponds to a subspace linked to at least one insight. The lowest tick (*) indicates no restriction on that field. 
Analysts can brush multiple non-adjacent axes to filter and highlight relevant subspaces, enabling efficient exploration of targeted insights and facilitating interactions with other views.
% \sout{for efficient exploration of targeted insights}

% \subsection{Insight Score View}
% % Similar to the Subspace Filtering View, 
% As introduced in Section~\ref{sec-insight-representation}, significance and impact are two major metrics that can delineate the importance of a data insight.
% When an analyst is facing a new dataset and has no detailed knowledge about the dataset, the importance and impact scores of insights can give them useful hints on where to start their initial exploration.
% Inspired by this, we design the insight score view to visualize the importance and impact of all the insights through a scatterplot.
% As shown in Fig.~\ref{fig_UI}D, the x-axis represents the significance of each insight and the y-axis indicates the impact of each insight.
% Each insight is represented as a gray dot in the scatterplot and their position explicitly indicates their impact and significance.
% By observing the dot distribution in the scatterplot, analysts can easily know the overall importance distribution of all the data insights automatically-extracted from the given dataset.

% % a quick overview of the insight distribution and their relative importance. 

% Interactions are also enabled in the Insight Score View. Analysts can select a set of insights by brushing on the scatterplot. Insights out of the brush will be assigned a low opacity in the Insight Map View for further exploration. 
% % \yong{Yong, I have changed this part}
\stitle{Insight Score View} In the \textit{insight score view}, we designed a scatterplot to display the two core attributes of insights: significance and impact, as introduced in Section~\ref{sec-insight-representation}.
As shown in Fig.~\ref{fig_UI}D, the x-axis represents significance, and the y-axis represents impact. Each insight is shown as a gray dot, with its position reflecting both metrics. 
Analysts can easily assess the overall importance distribution of insights by observing the scatterplot. 
Additionally, interactions allow analysts to select a set of insights by brushing, enabling exploration of only the selected insights for further analysis.

\subsection{Insight Map View}
% The Insight Map View aims to provide analysts with a concrete understanding of insight distribution, which is achieved through visualizing the details of each individual insight and displaying the correspondence between insights and subspaces and the similarity between different insights.
% The Insight Map View is the core visualization design of \sysNameWithoutSpace.
% As shown in Fig.~\ref{fig_UI}E, the insight map view consists of a map visualizing the individual insights and the subspace list displaying the detailed value range of each data field of the insights.
The \textit{Insight Map View}, the core visualization of \sysNameWithoutSpace, enables analysts to explore insight distribution, subspace relationships, and similarities. 
As shown in Fig.~\ref{fig_UI}E, it includes a map of individual insights along with a subspace list displaying value ranges for each data field.

\subsubsection{Insight Map}
% We use the metaphor of map to show the distribution of insights as well as a straightforward overview of each insight.
% Each insight is encoded as a glyph in the plane and the distance between insight glyphs indicates the similarity between any two insights.
We use a map metaphor to display insight distribution and provide an overview of each insight. 
Each insight is encoded as a glyph, with the distance between glyphs representing their similarity.

\begin{figure}[h!]
    \centering
    \begin{minipage}[b]{0.45\linewidth}
        \raggedleft
        \includegraphics[width=0.55\linewidth]{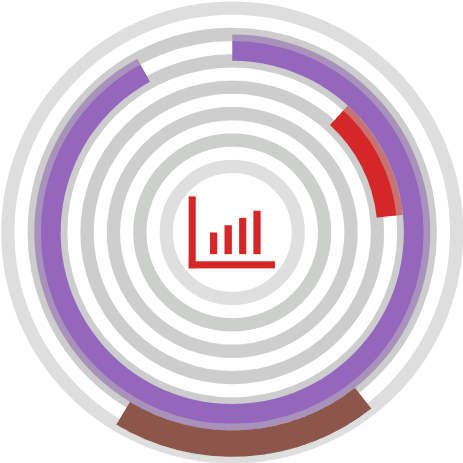} 
    \end{minipage}
    \hspace{0.05\linewidth} 
    \begin{minipage}[b]{0.45\linewidth}
        \raggedright
        \includegraphics[width=0.55\linewidth]{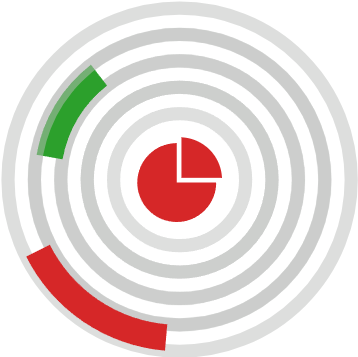} 
    \end{minipage}
    \caption{The glyph design for individual insights.}
    \label{fig_glyph}
\vspace{-3mm}
\end{figure}

\stitle{Insight Glyph}
Each insight is represented by five key elements (Section~\ref{sec-insight-representation}), but we focus on visualizing  \textit{type}, \textit{subspace}, and \textit{breakdown}, as insight scores are shown separately in the \textit{Insight Score View} and measure depends on the insight type.
As shown in Fig.\ref{fig_glyph}, concentric circles represent data fields, with arcs encoding value ranges. 
Unspecified fields appear gray. 
The center icon indicates the insight type using symbols (e.g., bars for Trend, a pie chart for Attribution), while its color reflects the breakdown.
For example, in Fig.\ref{fig_glyph}a, a Top-1 insight has a red icon and arc, linking the breakdown to the third circle.
To manage visual clutter, only insights with scores above a user-configurable threshold are displayed as glyphs (Fig.~\ref{fig_UI}E$_2$), whereas lower-scoring insights appear as dots to reduce occlusion.

% \yong{Q1. What does the icon color represent? Q2. The scalability issues of the glyph can be arguing through 1) interactions in the data distribution view; 2) most of the datasets do not have attributes of >= 5 data fields.}

% \textbf{Insight Layout}:
% To provide an overview based on the insight similarity, we project all insights onto the 2D canvas by dimension reduction techniques which takes the insight embedding as the input vector. Users are allowed to select embedding methods discussed above according to their exploration preference. We provides five data reduction techniques for users to select, including isometric mapping ~\cite{tenenbaum2000global}, multi-dimensional scaling (MDS)~\cite{borg2005modern}, spectral embedding~\cite{ng2001spectral}, and t-distributed stochastic neighbor embedding {(t-SNE)}~\cite{van2014accelerating}. Thus, the distance between the two insights on the canvas reflects the similarity of the insight. We choose the t-SNE method as the default option.  

% Links between different insights.
% To further inform analysts of the possible similarity between insights, {\sysNameWithoutSpace} also supports an explicit highlighting of insights sharing the same \textit{breakdown values}. Fig.~\ref{fig_UI}E$_2$ shows an example, where all the insights sharing the same \textit{breakdown value} with the current insight are connected to the current insight glyph with gray curves when the analyst clicks the current insight glyph.
\stitle{Insight Layout}
To provide an overview based on insight similarity, we project all insights onto a 2D canvas using dimensionality reduction techniques, with the insight embedding as the input vector. 
Users can choose from five methods, including isometric mapping~\cite{tenenbaum2000global}, multi-dimensional scaling (MDS)~\cite{borg2005modern}, spectral embedding~\cite{ng2001spectral}, and t-SNE~\cite{van2014accelerating}, based on their preferences. The distance between insights on the canvas reflects their similarity, with t-SNE set as the default method.
To further highlight insight similarities, {\sysNameWithoutSpace} allows users to visually connect insights with the same \textit{breakdown values}. Fig.~\ref{fig_UI}E$_2$ shows an example where insights with the same \textit{breakdown value} as the selected one are linked together with gray curves.

\stitle{Insight Density}
After showing all the insights in the \textit{Insight Map View}, we further employ the kernel density estimation (KDE) approach~\cite{turlach1993bandwidth} to estimate the density of data insight distributions in the \textit{Insight Map View} (Fig.~\ref{fig_UI}E). Then, we further visualize the insight density by drawing density contours. 
The higher density of insight distribution is visualized by the contours and the darker gray colors, providing analysts with a quick overview of the overall distribution of different insights.

% \subsubsection{Subspace List}
% The subspace list presents all the subspaces that correspond to at least one automated insight.
% As shown in Fig.~\ref{fig_UI}E$_1$, each subspace is represented by a row in the subspace list and each column is a data field.
% For a specific subspace, its value range of each data field is represented by the yellow bar on that cell, where the maximum and minimum values of that cell are determined by the maximum and minimum values of the whole dataset in that data field.
% By viewing the subspace list, analysts can easily identify the distribution of data subspaces and the overall similarity between different subspaces.
% The subspaces in the list are sorted by the number of data insights that are extracted from it, which can help analysts focus on those subspaces that are relatively important in terms of the number of insights.
% By clicking a subspace, all the corresponding insights on the map will be highlighted and linked to this subspace, enabling an easy track of related insights.
\subsubsection{Subspace List}
% The subspace list shows all subspaces corresponding to at least one automated insight. 
% As shown in Fig.~\ref{fig_UI}E$_1$, each row represents a subspace, and each column corresponds to a data field. 
% For a specific subspace, the value range for each field is represented by a yellow bar in the cell, with the range determined by the dataset's maximum and minimum values for that field. 
% By viewing the subspace list, analysts can easily identify the distribution of data subspaces and the overall similarity between different subspaces.
% The subspaces in the list are sorted by the number of insights extracted from them, guiding analysts to focus on those subspaces that are relatively important in terms of the number of insights.
% Clicking a subspace highlights all corresponding insights on the map, linking them for easy exploration.
The Subspace List displays all subspaces linked to at least one automated insight (Fig.~\ref{fig_UI}E$_1$). 
Each row represents a subspace, with columns corresponding to data fields. 
Yellow bars indicate value ranges, scaled to the dataset’s min-max values. 
Subspaces are sorted by the number of extracted insights, helping analysts focus on the most relevant ones. 
Clicking a subspace highlights its corresponding insights on the map, facilitating exploration.

\subsection{Individual Insight View}
The \textit{Individual Insight View} shows the details of each data insight. Specifically, when the users click on an insight glyph or an insight dot in the \textit{Insight Map View}, the corresponding insight details will be shown in the \textit{Individual Insight View}.
Following the practice of prior studies \cite{ding2019quickinsights, shi2020calliope}, each insight is illustrated with a representative chart.
By default, \textit{Trend} and \textit{correlation} insights are visualized using line charts, while \textit{Attribution} insights are shown with pie charts, as it aligns with human perception of proportions. Additionally, {\sysNameWithoutSpace} allows users to modify the visualization of insights as needed based on their preferences.

Also, to further clarify the insight details, when the users hover a mouse over the chart, a brief text description for the insight will be shown as well, which explicitly explains more details of the insight.
% For example,
% Fig.~\ref{fig-case-study-carsales}i shows an insight of \textit{Change Point}, which is clarified as ``the decreasing car sales of Volkswagen has a change point in 2009 and the sales of Volkswagen start to grow up after 2010''.
% The text description is based on templates for different types of insights.
% Here, we mainly adopt the text templates introduced in the prior study~\cite{shi2020calliope}.

\subsection{Interaction Design}
{\sysNameWithoutSpace} offers interactive exploration of automated insights. For instance, analysts can customize settings via the control panel and filter data fields by clicking column names (Fig.\ref{fig_UI}B). 
Multi-criteria filtering allows refinement through brushing axes (Fig.\ref{fig_UI}E$_1$), adjusting scores (Fig.\ref{fig_UI}D$)$, or selecting insight types (Fig.\ref{fig_UI}E$_3$). 
Clicking an insight glyph highlights related insights (Fig.\ref{fig_UI}E$_2$), while zooming, hovering, or double-clicking reveals detailed descriptions (Fig.\ref{fig_UI}b).

\renewcommand{\arraystretch}{1.2}
\begin{table*}
  \centering
  \resizebox{0.9\textwidth}{!}{
    \begin{tabular}{  m{1.5em} | m{55em} }
    \hline
    Q1 & Overall, will the proposed visual analytics system be useful for you to explore different datasets and insights? Why? \\
    \hline
    Q2 & Is the proposed system easy/difficult for you to gain an overview of the data distribution (the Data Distribution View)? Why? \\
    \cline{2-2}
    Q3 & Is the proposed system easy/difficult for you to gain an overview of the insight distribution (the Insight Score View and the Insight Map View)? Why? \\
    \cline{2-2}
    Q4 & Is the proposed system easy/difficult for you to explore the relation between different insights, or find new insights of your interest (the Insight Map View)? Why? \\
    \cline{2-2}
    Q5 & Is it easy/difficult for you to check the detailed individual insights (the Subspace Filtering view and Individual Insight View)? Why? \\
    \hline
    Q6 & Overall, is it easy/difficult for you to understand the visual designs of the proposed system? \\
    \cline{2-2}
    Q7 & Overall, is it convenient for you to interact with the system? \\
    \hline
    Q8 & What are the major pros and cons of the proposed approach (especially compared with the existing methods you used for exploring data insights)? \\
    \hline
  \end{tabular}}
  \vspace{3mm}
  \caption{The questionnaire is composed of four parts: usefulness (Q1), effectiveness (Q2 - Q5), usability (Q6 - Q7) and pros and cons (Q8).}
  \label{table-interview}
  \vspace{-5mm}
\end{table*}

\section{Case Study}
\label{sec-case-study}

% In this section, we conducted two case studies to demonstrate the effectiveness of {\sysNameWithoutSpace}.
% Two different real datasets are used in the case studies: \textit{car sales} and \textit{NBA statistics}.
% The \textit{car sales} dataset~\footnote{\url{https://www.microsoft.com/en-us/research/project/quickinsights/}} records the sales of different car from 2007 to 2011, which consists of 275 rows and 5 data fields.
% The \textit{NBA statistics} dataset~\footnote{\url{https://www.nba.com/}} records the detailed performance statistics of different NBA players during different seasons, which consists of 20,973 rows and 38 data fields.

% \sout{We present a case study to evaluate the effectiveness of {\sysNameWithoutSpace} using the real-world \textit{NBA Statistics} dataset~\footnote{\url{https://www.nba.com/}}, which contains detailed performance metrics of NBA players across multiple seasons. The dataset comprises 20,973 records and 38 data fields.}

We present a case study to evaluate the effectiveness of {\sysNameWithoutSpace} using a real-world dataset, the NBA Statistics~\footnote{\url{https://www.nba.com/}}, which is crawled from the official NBA website. The dataset records detailed performance statistics of different NBA players across multiple seasons, comprising 20,973 records and 38 data fields.

In this case study, an NBA fan used {\sysNameWithoutSpace} to analyze the NBA dataset and explore the correlation between players' performance and their roles. 
Fig.~\ref{fig_UI} shows part of his exploration. After loading the dataset, he selected all data field dimensions (Fig.~\ref{fig_UI}A) to conduct a comprehensive exploration of all the data insights.
He quickly noticed dense data insights with high significance
and impact at the top of the Insight Map View (Fig.~\ref{fig_UI}E), indicated by darker contours and glyphs.

The insights with low significance and impact are displayed as dots instead of glyphs. Therefore, the user clicked an insight glyph in this area, revealing insight details in the Individual Insight View (Fig.~\ref{fig_UI}d). The chart showed the proportion of the total assists for five different NBA player roles (i.e., Point Guard, Shooting Guard, Small Forward, Power Forward and Center). 
As shown in Fig.~\ref{fig_UI}e, the pie chart shows an almost even distribution of points scored by NBA players across five roles. 
This observation contradicted his expectation, as he believed certain roles should have greater scoring responsibility.
In his view, \textit{Center} scored more in the 1960s–70s, while \textit{Small Forwards} took on a bigger scoring role in modern games~\footnote{\url{https://medium.com/@asktumma/the-evolution-of-the-nba-big-man-704184edefe5}}.
However, scoring data from 1947 to now (Fig. 2e) suggests that different player roles have contributed evenly to successful shots in terms of their average performances since 1947.
% \sout{and discovered that successful shots have been almost evenly distributed across all player roles since 1947 (Fig.~\ref{fig_UI}e). This revised his understanding, as he had previously assumed that certain roles, like \textit{Center} in the 1960s and 1970s or \textit{Small Forward} in modern games, would score more~\footnote{\url{https://medium.com/@asktumma/the-evolution-of-the-nba-big-man-704184edefe5}}. }

The above insights sparked the user’s interest in exploring the evolution of NBA players' scoring. He further focused on the \textit{Trend} and \textit{Change Point} insights (Fig.~\ref{fig_UI}F) by filtering. 
% \sout{A \textit{Change Point} insight stood out (Fig.~\ref{fig_UI}f), showing a sudden jump in total scores in 1968, which intrigued him. }
He easily identified an interesting \textit{Change Point} insight, as shown in Fig.~\ref{fig_UI}f. It visualized the total scores of all NBA players each year since 1947 and highlighting a noticeable \textit{Change Point} in 1968, i.e., a sudden jump in total scores. This unexpected shift immediately caught his interest, prompting him to explore its possible causes.

The user then checked \textit{Attribution} insights again to identify key contributors to the scores.
Specifically, he examined the score proportions across different leagues, as shown in Fig.~\ref{fig_UI}a.
It was clear that NBA and ABA were the major leagues, with NBA contributing most of the scores. He further explored the \textit{Trend} insights for both leagues.
% \sout{noticing that NBA scores (Fig.~\ref{fig_UI}b) followed a smoother trend compared to the total scores (Fig.~\ref{fig_UI}f), especially around 1968. }
He found that NBA scores (Fig.~\ref{fig_UI}b) follow a similar overall trend to the combined NBA and ABA scores (Fig.~\ref{fig_UI}f), but NBA's score changes are much smoother around 1968.
He speculated that the abrupt change in 1968 was due to the establishment of the ABA league.
% \sout{which he confirmed through further research}
By conducting researching on the related
materials, his speculation was confirmed: the ABA league was established in 1967 and participated in games for the first time in 1968~\footnote{\url{https://zh.wikipedia.org/wiki/NBA}}.
The ABA's introduction led to the fast development of
basketball games and the \textit{Change Point} in Fig.~\ref{fig_UI}f.

\section{User Interviews}
\label{sec-user-interviews}

We conducted semi-structured interviews with four expert users who regularly perform data analysis and exploration to evaluate the usefulness and usability of {\sysNameWithoutSpace}.
We first describe the interview settings, including participants, tasks, and procedures, followed by a summary of user feedback.

\subsection{Participants}
Our interviews involved four expert users who regularly engage in data analysis and insight discovery as part of their their work and research. Three participants (\textit{E1}, \textit{E2}, and \textit{E3}) are analysts or researchers
with backgrounds in data mining and data analytics. 
\textit{E1} is a data analyst and engineer at a local company with 12 years of experience in business intelligence and data mining. 
\textit{E2} and \textit{E3} are researchers in databases and data mining at a local university, each with at least two years of experience in data analysis. 
The fourth participant (\textit{E4}) is a data visualization researcher at a local university with over three years of experience in data analytics and visualization. 
None of the participants had prior exposure to {\sysNameWithoutSpace} before the user interviews. We conducted the user interviews with each expert user individually.

\subsection{Task}
We asked the participants to conduct data analysis and insight exploration on real multidimensional datasets.
We asked the participants to fulfill two carefully-designed tasks to evaluate the effectiveness and usability of {\sysNameWithoutSpace} in facilitating insight exploration.
\squishlist
\item \textbf{T1. Insight Overview:} Investigate the overall distribution of data insights and data attributes.
\item \textbf{T2. Insight Detail Exploration:} Inspect the relations between insights and check the details of individual insights.
\squishend
\textbf{T1} aims to assess whether {\sysNameWithoutSpace} can effectively provide users with a quick overview of insight and data attribute distributions. 
For \textbf{T1}, the participants will mainly use the Data Distribution View, Subspace Filtering View, Insight Score View and Insight Map View.
\textbf{T2} is designed to evaluate whether {\sysNameWithoutSpace} can enable convenient exploration of the relation among different insights and as individual insight details, where the Insight Map View and the Individual Insight View will be main views to finish this task.

After they finished the two tasks, we further conducted a post-study questionnaire to collect their detailed feedback and suggestions on the workflow, visualization design, interaction and overall usability of {\sysNameWithoutSpace}. They were also encouraged to compare {\sysNameWithoutSpace} with their previous ways of exploring data insights of multidimensional datasets.
Table~\ref{table-interview} shows the detailed questions in our post-study questionnaire.
The whole user interview lasted about 60 minutes.

\subsection{Results}
% Overall, our in-depth user interviews demonstrated that {\sysNameWithoutSpace} can effectively help participants gain a quick overview and conduct detailed exploration of data insights on multidimensional datasets.
% We summarize the detailed feedback and comments by the participants during the interviews as follows:
Our user interviews confirmed that {\sysNameWithoutSpace} effectively enables quick overviews and in-depth exploration of data insights within multidimensional datasets. 
The key participant feedback is summarized below.

\stitle{Usefulness} Overall, all the participants confirmed the usefulness of {\sysNameWithoutSpace}.
\textit{E1} commented that ``{\sysNameWithoutSpace} is quite useful, as it can help me directly see the major key features of different insights such as their types and subspaces, and the visualizations, especially the Insight Map View, show the insight distribution and the relations among them in an intuitive and convenient manner, which is quite helpful for exploring similar data insights''.
% \textit{E2} said that the visualizations are straightforward and enable a quick and comprehensive understanding of the whole dataset and the underlying data insights, which is much more convenient than other methods that he often uses for data analysis and insight discoveries, such as programming or using some data analysis software like Excel and Tableau. 
% \textit{E3} and \textit{E4} pointed out that they have never seen such tools like {\sysNameWithoutSpace} that can support a comprehensive exploration of automated data insights, especially in terms of gaining a data insight overview and conveniently investigating similar insights.
\textit{E2} found the visualizations straightforward, providing a quick and comprehensive dataset understanding, surpassing traditional tools like Excel and Tableau. 
\textit{E3} and \textit{E4} noted that they had not encountered similar tools offering both an overview of automated insights and convenient exploration of related insights.

\stitle{Effectiveness}
All participants agreed that {\sysNameWithoutSpace} effectively supports exploring insight distributions, identifying relevant data insights, and examining details. \textit{E1} and \textit{E4} found the Insight Map View intuitive for inspecting distributions and discovering similar insights. \textit{E2} and \textit{E3} needed a brief adjustment period but ultimately appreciated its interactive features for linking related insights. 
% Additionally, all participants found the Individual Insight View useful for easily examining insights through clear charts and concise descriptions.

% All the participants agreed that {\sysNameWithoutSpace} can effectively assist them in investigating the overall insight distributions, finding the data insights of their interest and further checking the insight details. 
% For the Insight Map View, \textit{E1} and \textit{E4} mentioned that they can easily understand the overall visualization designs and use it to inspect the insight distribution and discover similar data insights.
% \textit{E2} and \textit{E3} said that it took them a few minutes to
% be familiar with the glyph designs and the layout of the insight glyphs in the Insight Map View. But after they understood the overall visual designs and enabled interactions, they appreciated the Insight Map View and confirmed its effectiveness in helping them find similar data insights through the interactive links between related insights as well as checking the overall insight information through the insight glyphs.
% Also, all the participants found the Individual Insight View useful and effective, as it enabled them to check the individual insights in detail easily with the straightforward charts and brief text descriptions.

\stitle{Usability} 
% The participants appreciated the well-designed visual analytics system and the convenient interactions supported in it.
% \textit{E1} praised the Insight Map View and said that he pretty likes the interactive links between the subspace list and the data insight glyphs as well as the links between the current insight glyph and other similar data insights, which can effectively help him with the linked analysis from data subspaces to data insights and further the similar data insights.
% \textit{E3} and \textit{E4} enjoyed the density contours and emphasized that it can help them gain a quick overview of insight distributions when there are a large number of data insights in the input dataset.
The participants appreciated the well-designed visual analytics system and its intuitive interactions. \textit{E1} particularly praised the Insight Map View, highlighting the interactive links between subspaces, insight glyphs, and similar insights. 
\textit{E3} and \textit{E4} also found the density contours useful for quickly grasping insight distributions in large datasets.

\stitle{Limitations and Suggestions} 
Despite the positive feedback, participants also identified several limitations of {\sysNameWithoutSpace} and offered suggestions for improvement. 
\textit{E1} noted that the automated insight extraction method (i.e., QuickInsights \cite{ding2019quickinsights}) used in {\sysNameWithoutSpace} lacks intuitiveness and explainability, which affects the clarity of the extracted insights. He also pointed out that some extracted insights are neither particularly interesting nor important, suggesting that more advanced automated extraction methods should be explored. 
\textit{E3} observed that when multiple insights are related to the current one, the links between them are overly dense, leading to visual clutter. 
\textit{E2} and \textit{E3} mentioned that comparing different data insights in the Insight Map View is not intuitive, and {\sysNameWithoutSpace} could better support explicit comparisons between different insights or groups of insights. 
\textit{E3} suggested that {\sysNameWithoutSpace} should allow users to interactively configure dataset subspaces for insight extraction instead of limiting them to exploring only precomputed insights defined by developers.

\section{Conclusion} % and Future Work
% In this paper, we proposed {\sysNameWithoutSpace}, an interactive visual analytics approach that combines automated insight extraction with interactive visual exploration to facilitate quick overview and in-depth exploration of automated data insights.
% Specifically, we presented a generic quantitative technique to characterize each data insight (i.e., data insight embedding) and further measure the similarity between data insights.
% Also, we introduced novel visualization designs using a metaphor of map to display the insight distribution as well as illustrate the relation (e.g., similarity) between data insights.
% We further conducted extensive evaluations on our approach through three case studies on real datasets and carefully-designed user interviews with expert users. 
% The evaluation results demonstrate the effectiveness and usability of our approach in assisting analysts with data insight exploration.
% The lessons we learned and possible limitations are also explicitly discussed.

We presented {\sysNameWithoutSpace}, an interactive visual analytics system that integrates automated insight extraction with interactive exploration. It enables analysts to efficiently obtain an overview and explore data insights through novel embedding-based similarity measurements and map-based visualization.
Evaluations through a case study and expert interviews demonstrate its effectiveness. While scalability and extraction quality still remain challenges, our design effectively mitigates these through filtering and future extensibility. {\sysNameWithoutSpace} lays a solid foundation for enhancing data-driven discovery, with potential extensions to support more insight types and relationships.
% \input{src/10-conclusion} Merged with discussion

% \begin{acks}
% \end{acks}

\bibliographystyle{ACM-Reference-Format}
\bibliography{refs}

\appendix

\end{document}